\documentclass[preprint, prl, amssymb,  aps, showpacs,preprintnumbers,
amsmath,showkeys,floatfix]{revtex4}

\usepackage{epstopdf}
\usepackage{graphics}
\usepackage{graphicx}
\usepackage{dcolumn}
\usepackage{bm}
\usepackage{longtable}
\usepackage{epsfig}
\usepackage{times}
\usepackage{url}
\usepackage{color}

\begin{document}
%%%%%%%%%%%%%%%%%%%%%%%%%%%%%%%%%%%%%%%%%%%%%%%%%%%%%%%%%%%%%%%%%%%%%%%%%%%%%%%%
%%%%%%%%%%%%%%%%%%%%%%%%%%%%%%%%%%%%%%%%%%%%%%%%%%%%%%%%%%%%%%%%%%%%%%%%%%%%%%%%
\title{Coherent storage and phase modulation of single hard x-ray photons using nuclear excitons}

\author{Wen-Te \surname{Liao}}
\email{Wen-Te.Liao@mpi-hd.mpg.de}

\author{Adriana \surname{P\'alffy}}
\email{Palffy@mpi-hd.mpg.de}

\author{Christoph~H. \surname{Keitel}}
%\email{Keitel@mpi-hd.mpg.de}

\affiliation{Max-Planck-Institut f\"ur Kernphysik, Saupfercheckweg 1, D-69117 Heidelberg, Germany}
\date{\today}
%%%%%%%%%%%%%%%%%%%%%%%%%%%%%%%%%%%%%%%%%%%%%%%%%%%%%%%%%%%%%%%%%%%%%%%%%%%%%%%%
%%%%%%%%%%%%%%%%%%%%%%%%%%%%%%%%%%%%%%%%%%%%%%%%%%%%%%%%%%%%%%%%%%%%%%%%%%%%%%%%
\begin{abstract}
Coherent storage and phase modulation of x-ray single-photon wave packets  in resonant scattering of light off nuclei is investigated theoretically. 
We show that by switching off and on again the magnetic field in the nuclear sample, phase-sensitive storage of photons in the keV regime can be achieved. Corresponding
$\pi$ phase modulation of the stored photon can be accomplished if the retrieving magnetic field is rotated by $180^{\circ}$. The development of such 
x-ray single-photon control techniques is a first step towards forwarding quantum optics and quantum information to shorter wavelengths and more compact photonic devices.

\end{abstract}
%%%%%%%%%%%%%%%%%%%%%%%%%%%%%%%%%%%%%%%%%%%%%%%%%%%%%%%%%%%%%%%%%%%%%%%%%%%%%%%%
%%%%%%%%%%%%%%%%%%%%%%%%%%%%%%%%%%%%%%%%%%%%%%%%%%%%%%%%%%%%%%%%%%%%%%%%%%%%%%%%
\maketitle
%%%%%%%%%%%%%%%%%%%%%%%%%%%%%%%%%%%%%%%%%%%%%%%%%%%%%%%%%%%%%%%%%%%%%%%%%%%%%%%%
%-----------Text body-----------------------------------------------------------

Seeking for versatile solutions for quantum and classical computing on the most compact scale is 
one of the crucial objectives in both fundamental physics and information technology.
The photon as  flying qubit is anticipated to be the fastest information carrier 
and to provide the most efficient computing implementation. However, extending Moore's law \cite{Moore1998}
to the future quantum photonic circuits must meet the bottleneck of the diffraction limit, 
i.e., few hundred nm for the optical region. Forwarding optics and quantum information to shorter wavelengths in the x-ray region has the potential of shrinking computing elements in  future photonic devices such as the quantum photonic  circuit \cite{Politi2008}. This is strongly related to the development and availability of compact x-ray sources based on table-top  plasma wigglers \cite{kneip2010bright} and magnet undulators \cite{fuchs2009laser} or x-ray high-harmonic generation with optical coherent light sources \cite{chen2010bright}. The realization of a short wavelength quantum photonic circuit requires mastery of x-ray optics and powerful control tools of single-photon wave packet amplitude, frequency, polarization and phase \cite{Specht2009}. The development of x-ray optics elements has made already significant progress  with the realization of x-ray diamond mirrors \cite{Shvydko2004,Shvydko2010,Shvydko2011} and cavities \cite{Ishikawa2008},  hard x-ray waveguides \cite{PfeifferWGuide,JarreWGuide} and the Fabry-P\'erot resonator \cite{Liss2000,Shvydko2003,Ishikawa2005}. Efficient coherent photon storage for photon delay lines and x-ray phase modulation, preferably even for single-photon wave packets, are next milestones yet to be reached.

Moving towards the interactions in the x-ray  regime \cite{Buth2007,Zepf2009,Schafer2009,Shwartz2011,Young2011,Rohringer2012}, also new physical systems come into play, e.g.,   nuclei with low-lying collective states naturally arise as  candidates for x-ray quantum optics studies. Nuclear quantum optics \cite{Kocharovskaya1999,Coussement2002,Buervenich2006} and nuclear coherent population transfer \cite{Liao2011} are rendered experimentally possible by 
the advent and commissioning of x-ray free electron lasers (XFEL) \cite{slac,Sacla,xfel}. Coherent control tools based on nuclear cooperative effects \cite{Van1999, shvydko2000, Roehlsberger2004, Roehlsberger2010,Roehlsberger2012} are known also from nuclear forward scattering (NFS) experiments with third-generation synchrotron light sources. The underlying physics here relies on the delocalized nature of the nuclear excitation produced by coherent XFEL or synchrotron radiation (SR) light, i.e., the formation of  so-called nuclear excitons. 
Key examples in this direction is how manipulation of the hyperfine magnetic field in NFS systems  provides means to store nuclear excitation energy \cite{Shvydko1996} and in turn to generate  keV single-photon entanglement \cite{Palffy2009}. 

%%%%%%%%%%%%%%%%%%%%%%%%%%%%%%%%%%%%%%%%%%%%%%%%%%%%%%%%%%%%%%%%%%%%%%%%%
\begin{figure}[b]
\vspace{-0.4cm}
  \includegraphics[width=8.5cm]{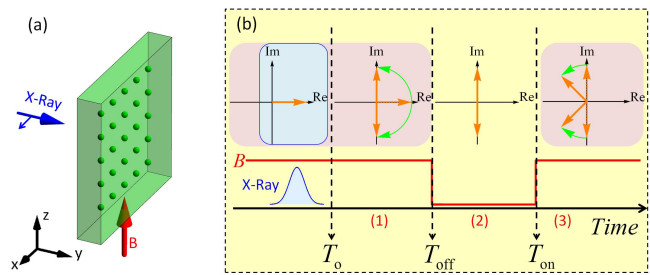}
  \caption{\label{fig1}
  (a) NFS setup. The blue arrow depicts the linear polarized x-ray pulse propagating in $y$  direction and {\bf B} is the external magnetic field  initially parallel to the $z$-axis. (b) The time dependence of the hyperfine magnetic field {\bf B} (red line) and the corresponding dynamics of the nuclear currents (orange arrows). The dynamics will be surveyed in three temporal domains: (1) $T_{\mathrm{o}}<t<T_{\mathrm{off}}$; (2) $T_{\mathrm{off}}<t<T_{\mathrm{on}}$; (3) $T_{\mathrm{on}}<t$.
  }
\end{figure}
%%%%%%%%%%%%%%%%%%%%%%%%%%%%%%%%%%%%%%%%%%%%%%%%%%%%%%%%%%%%%%%%%%%%%%%%% 

In this Letter, we  present two important control tools for single hard x-ray photons using resonant scattering of light off nuclei in a NFS setup. The formation of a nuclear exciton consisting of a single delocalized excitation  opens the possibility to control the coherent decay and therefore emission of the scattered photon.  Making use of this feature, we first put forward how to coherently store  a single hard x-ray photon for time intervals of 10-100 ns by turning off the hyperfine magnetic field in a NFS system. The stored single photon can be released by turning on the magnetic field. We emphasize that our scheme  conserves not only the excitation energy, as already pioneeringly demonstrated in Ref.~\cite{Shvydko1996}, but also the photonic polarization and phase beyond the ps time range.  
Next, we show how to modulate the stored photon with a phase shift of $\pi$  by using a releasing hyperfine magnetic  field oriented in the opposite direction to the initial one.
For the measurement of this $\pi$-phase shift of the retrieved photon, we refer to the echo technique using two nuclear targets \cite{Smirnov1996NE, jex1997, Smirnov2005} and 
demonstrate for the first time a magnetically induced nuclear exciton echo without any mechanical vibration of the targets. This feasible echo two-sample setup can also be used for phase-sensitive photon storage  involving a mere rotation of the hyperfine magnetic field by 180$^{\circ}$.

The typical NFS setup involves a solid-state target containing $^{57}$Fe. A x-ray pulse with meV  bandwidth (either SR or coherent XFEL light) tuned on the 14.413 keV nuclear transition from the ground state to the first excited state shines perpendicular to the nuclear sample, as shown in Fig.~\ref{fig1}(a).  SR typically produces at most one excited nucleus per pulse, thus providing a reliable single-excitation and single released photon scenario. The disadvantage here is that the initial photonic phase is undefined. Coherent x-ray light from seeded or oscillator XFEL \cite{sxfel.Feldhaus,sxfel.Saldin,XFELO} with a well-defined photonic phase can be used at low intensities such as to keep the excitation rate below one nucleus per pulse in the sample and guarantee single photons. Control over the number of excited nuclei per pulse can be achieved either by using x-ray partial reflection or partial transmission on silicon mirrors \cite{Shvydko2004} in order to limit the laser beam intensity or by varying the concentration of $^{57}$Fe nuclei in the target. An externally applied magnetic field {\bf B} parallel to the $z$ axis induces the nuclear hyperfine splitting of the ground and excited $^{57}$Fe nuclear states of spins $I_g$=1/2 and $I_e$=3/2, respectively. Depending on the pulse polarization, different hyperfine transitions will be driven. In  the following we consider the x-ray field linearly polarized parallel to  the $x$ axis driving the two  $\Delta m=m_{e}-m_{g}=0$ magnetic dipole transitions, where  $m_{e}$ and $m_{g}$ denote  the projections of the excited and ground state nuclear spins on the quantization axis, respectively.

The dynamics of the density matrix $\widehat{\rho}$ is governed by the Maxwell-Bloch equations \cite{Crisp1970, Shvydko1999N, Scully2006, Palffy2008}: 
\begin{eqnarray}
&&
\partial_{t}\widehat{\rho} = \frac{1}{i\hbar}\left[ \widehat{H},\widehat{\rho}\right]+\widehat{\rho}_{s}\, ,
\nonumber\\
&&
\frac{1}{c}\partial_{t}\Omega_{}+\partial_{y}\Omega_{}=i\eta\left(a_{31}\rho_{31}+a_{42}\rho_{42}\right)\, ,
\label{eq1}
\end{eqnarray}
with the interaction Hamiltonian
\begin{equation}
\widehat{H} = -\frac{\hbar}{2}\left( 
\begin{array}{cccc}
  2\Delta_{g} & 0 & a_{13}\Omega^{*}_{} & 0\\
  0 & -2\Delta_{g} & 0 & a_{24}\Omega^{*}_{}\\
  a_{31}\Omega_{} & 0 & -2(\varDelta+\varDelta_{e}) & 0\\
  0 & a_{42}\Omega_{} & 0 & -2(\varDelta-\varDelta_{e})
\end{array}  
\right)\, .
\nonumber
\end{equation}
In the equations above  $\varDelta$ is the x-ray detuning to the 14.4 keV transition assumed to be zero and $\varDelta_{g(e)}$ denotes the Zeeman energy splitting of the nuclear ground (excited) state  proportional to the magnetic field {\bf B}. In Eq.~(\ref{eq1}), $\rho_{eg}=A_{e}A^{*}_{g}$ for $e\in \{1,2\}$ and $g\in \{3,4\}$ are the density matrix elements of $\widehat{\rho}$ for the nuclear wave function $|\psi\rangle= A_{1}|1\rangle+A_{2}|2\rangle+A_{3}|3\rangle+A_{4}|4\rangle$. The ket vectors are the eigenvectors of the two ground and two excited states hyperfine levels with $m_g=-1/2$, $m_g=1/2$, $m_e=-1/2$ and $m_e=1/2$, respectively. Furthermore,  $a_{eg}=a_{ge}=\sqrt{2/3}$ are the corresponding Clebsch-Gordan coefficients \cite{Shvydko1998, Palffy2008} for the  $\Delta m=0$ transitions and $\widehat{\rho}_{s}$ describes the spontaneous decay \cite{Scully2006}.  The parameter $\eta$ is defined as $\eta=\frac{6\Gamma}{L}\alpha$, where $\Gamma=1/141.1$ GHz is the spontaneous decay rate of excited states, $\alpha$ represents the effective resonant thickness \cite{Crisp1970, Shvydko1998, Shvydko1999N} and $L=10$ $\mu$m  the thickness of the target, respectively. Further notations are $\Omega_{}$ for the Rabi frequency which is propotional to the electric field $\vec{E}$ of the x-ray pulse  \cite{Scully2006, Palffy2008} and $c$ the speed of light.

%%%%%%%%%%%%%%%%%%%%%%%%%%%%%%%%%%%%%%%%%%%%%%%%%%%%%%%%%%%%%%%%%%%%%%%%%
\begin{figure}[b]
\vspace{-0.6cm}
  \includegraphics[width=8.5cm]{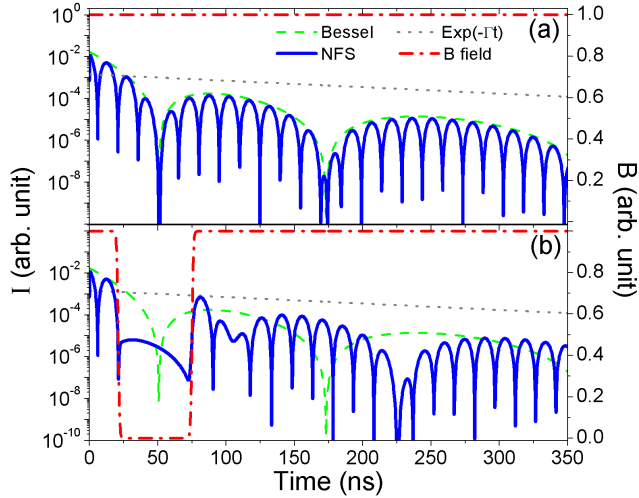}

 \caption{(a) Unperturbed NFS time spectrum: blue solid lines are the intensities of the  NFS signal, red dashed dotted lines denote qualitatively the applied magnetic field {\bf B}, the gray dotted lines are proportional to $e^{-\Gamma t}$  and the green dashed lines illustrate the dynamical beat \cite{Crisp1970, Shvydko1999H}. (b)
The hyperfine magnetic field is turned off at $t=21$ ns and turned back on at  $t=75$ ns.
  }\label{fig3}
\end{figure}
%%%%%%%%%%%%%%%%%%%%%%%%%%%%%%%%%%%%%%%%%%%%%%%%%%%%%%%%%%%%%%%%%%%%%%%%%
Fig.~\ref{fig1}(b) illustrates the time evolution of our  photon storage scheme. The external magnetic field {\bf B}, depicted by the red line,  is present before the x-ray pulse   impinges on the target at $T_{\mathrm{o}}$. At $T_{\mathrm{off}}$ the {\bf B} field is turned off and later turned back on at $T_{\mathrm{on}}$. The orange arrows   depict the time evolution of the nuclear transition current matrix elements as defined in Ref.~\cite{Shvydko1996}. In our treatment, this is equivalent with investigating the  coherence terms $i\rho_{42}$ and $i\rho_{31}$ \cite{Crisp1970, Shvydko1999N}. 

Initially, the ensemble of $^{57}$Fe nuclei is excited by the x-ray pulse  at $T_{\mathrm{o}}$. Subsequently, the purely real currents are abruptly built. In the time interval (1), the two currents start to rotate in  opposite directions on the complex plane with the factor of $e^{\pm i\bigtriangleup_{B}t}$ caused by the magnetic  field {\bf } until $t=T_{\mathrm{off}}$ when {\bf B} is turned off. The corresponding phase gain is $\pm \bigtriangleup_{B}\tau$. Here and in the following we have used for simplicity the notations $\bigtriangleup_{B}=\varDelta_{g}+\varDelta_{e}$ and $\tau=T_{\mathrm{off}}-T_{\mathrm{o}}$. Within the time interval (2), the quantum beat (arising from the interference between the two $\Delta m=0$ transitions) is frozen with the factor of $e^{\pm i\bigtriangleup_{B}\tau}$ since the hyperfine field has vanished, and only the dynamical beat  \cite{Crisp1970, Shvydko1998, Van1999} due to interference between multiple scattering processes  in the sample persists. During the time interval (3), the presence of the magnetic field makes  the quantum beat emerge again.

\begin{figure}[b]
\vspace{-0.6 cm}
  \includegraphics[width=8.5cm]{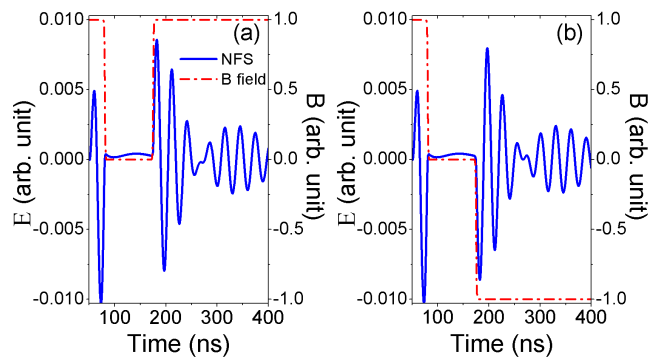}

 \caption{ Phase modulation of retrieved x-ray via reversing the applied magnetic field {\bf B}. Blue solid lines are the electric field of NFS signal, red dashed lines denote the applied magnetic fields {\bf B}. The {\bf B} field is turned off at $T_{\mathrm{off}}=80.5$ ns and then switched on at $T_{\mathrm{on}}=175$ ns, such that  (a) {\bf B}($t<T_{\mathrm{off}}$) = {\bf B}($t>T_{\mathrm{on}}$)  and  (b) {\bf B}($t<T_{\mathrm{off}}$) = -{\bf B}($t>T_{\mathrm{on}}$). 
  }\label{fig4}
\end{figure}

We numerically solve Eq.~(\ref{eq1}) with $\alpha=10$ and $\bigtriangleup_{B}=15\Gamma$, and present our  results in Figs.~\ref{fig3} and ~\ref{fig4}. 
The NFS signal intensities  $|\vec{E}_{}(t,L)|^{2}$ are compared with the spontaneous decay curves $e^{-\Gamma t}$ and the pure dynamical beat (for the case of no hyperfine splitting) $\left( \frac{\alpha}{\sqrt{\alpha\Gamma t}}J_{1}\left[2\sqrt{\alpha\Gamma t}\right]\right)^{2} e^{-\Gamma t}$  \cite{Crisp1970, Shvydko1999H}, where $J_{1}$ is the Bessel function of first kind. 
Fig.~\ref{fig3}(a) shows the unperturbed NFS time spectrum where both quantum beat and dynamical beat are observed. 
In Fig.~\ref{fig3}(b)  we demonstrate photon storage  by turning off the magnetic  field at $t=21$ ns (corresponding to a quantum beat minimum, $\bigtriangleup_{B}\tau=\pm N\frac{\pi}{2}$ with $N$ odd).  Both nuclear currents corresponding to the $\Delta m=0$ transitions  are frozen on the imaginary axis (see Fig.~\ref{fig1}(b)) and undergo destructive interference. In this case the intensity of the emitted radiation is  suppressed by three orders of magnitude. Later on, by turning the hyperfine magnetic field on again at $t=75$ ns, the unsuppressed photon signal is observed again within the time interval (3). Fig.~\ref{fig3} also shows that the stored nuclear excitation energy experiences spontaneous decay  during the storage time \cite{Shvydko1996}. 

The electric field envelopes of the scattered photon are presented in Fig.~\ref{fig4}.  In Fig.~\ref{fig4}(a), the magnetic field before $T_{\mathrm{off}}=80.5$ ns and that after $T_{\mathrm{on}}=175$ ns are the same and the phase before storage and after retrieving is continuous. If, however,  the retrieving magnetic field is applied in opposite direction as shown in Fig.~\ref{fig4}(b), the phase of the released photonic wave packet will be  modulated with a shift of $\pi$. This is caused by the effect of reversed time related with the change of sign of the hyperfine magnetic field \cite{Shvydko1994P, Shvydko1995}, i.e., all the nuclear currents evolve backwards in time.
Our density matrix calculations have been double-checked by the comparison with results from the iterative solution of the wave equations developed in Ref. \cite{Shvydko1996}. The agreement is complete for both electric field envelope and scattered light intensity, proving the equivalence of the two methods.

The most significant advantage of our scheme  is the conservation of the photonic polarization and phase. Storage of nuclear excitation energy  by magnetic field rotations in NFS experiments with SR was presented in Ref.~\cite{Shvydko1996}. This pioneering work has opened the avenue of coherent control applications with nuclei using magnetic switching.
However, the scheme in Ref.~\cite{Shvydko1996} is not phase-sensitive. 
Since  the magnetic Hamiltonian is not zero during the storage, both the polarization \cite{Palffy2010} and the phase of the particular polarization components cannot be stored and the properties of the released photon depend on the switching instants. With the advent of coherent XFEL sources and x-ray quantum optics and quantum information experiments, phase storage and modulation become crucial for many applications. So far, coherent trapping of hard x-rays in crystal cavities provides photon storage for time intervals in the ps range \cite{Ishikawa2008}. Our scheme provides robust phase and polarization storage of the x-ray photon on the 10-100 ns scale determined by the nuclear lifetime. 

%%%%%%%%%%%%%%%%%%%%%%%%%%%%%%%%%%%%%%%%%%%%%%%%%%%%%%%%%%%%%%%%%%%%%%%%%
%%%%%%%%%%%%%%%%%%%%%%%%%%%%%%%%%%%%%%%%%%%%%%%%%%%%%%%%%%%%%%%%%%%%%%%%%

In order to implement our phase-sensitive storage scheme experimentally, a material  with no intrinsic nuclear Zeeman splitting like stainless steel Fe$_{55}$Cr$_{25}$Ni$_{20}$ \cite{Smirnov1996NE, jex1997}   is required. The remaining challenge is to turn off and on the external magnetic fields of few Tesla on the ns time scale. According to our calculations for the case of Fig.~\ref{fig3}, the raising time of the {\bf B} field should be shorter than 50 ns (the raising time was considered 4 ns for all presented cases). This could be achieved by using small single- or few-turn coils and a moderate pulse current of approx. 15 kA from low-inductive high-voltage ``snapper'' capacitors \cite{Miura2003}. Another mechanical solution, e.g., the lighthouse setup \cite{Roehlsberger2000} could be used to move the excited target out of and into a region with confined static {\bf B} field. The sample is first excited while located in a first confined static  magnetic field  region. A fast rotation moves the sample out of this magnetic field region, and later on brings it under the action of a second static magnetic field. Simple geometrical considerations show that a displacement of the size of the sample thickness (about 3.5 $\mu$m) corresponds to a time interval of 10 ns at a rotation frequency of 70 kHz and rotor radius of 5 mm \cite{Roehlsberger2000}. The sample can be thus rotated out of the confined magnetic field region fast enough to provide switching times on the order of 10 ns.

%%%%%%%%%%%%%%%%%%%%%%%%%%%%%%%%%%%%%%%%%%%%%%%%%%%%%%%%%%%%%%%%%%%%%%%%%
\begin{figure}[t]
\vspace{-0.6 cm}
  \includegraphics[width=8.5cm]{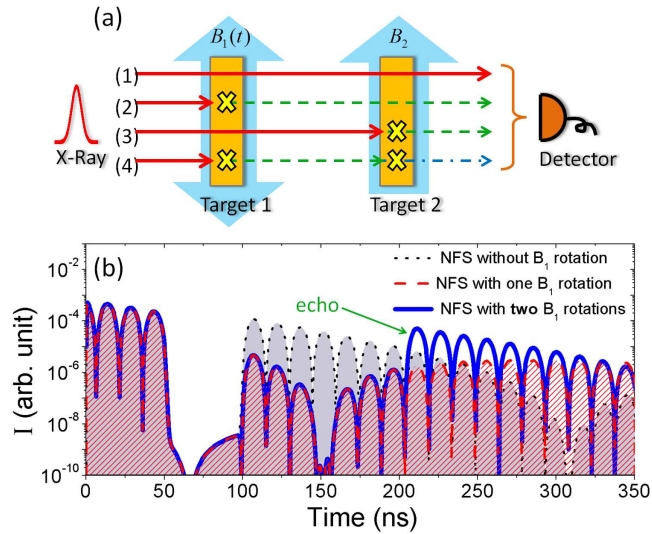}

 \caption{(a) Magnetically induced nuclear exciton echo setup with two targets. Yellow crosses illustrate the formation of the nuclear exciton. The light blue up-down thick arrows show the applied magnetic fields: the dynamical ${\bf B}_{1}(t)$ is applied to target 1, whereas the static ${\bf B}_{2}$ is  applied to target 2.
(b) NFS time spectrum for $\alpha_{1}=\alpha_{2}=1$ and $\vert\bigtriangleup_{B1}\vert=\bigtriangleup_{B2}=15\Gamma$. The magnetic field ${\bf B}_{1}(t)$ is turned off  at $T_{\mathrm{off}}=51$ ns and on at $T_{\mathrm{on}}=100$ ns. Black dotted line: ${\bf B}_{1}(t>T_{\mathrm{on}}) = {\bf B}_{1}(t<T_{\mathrm{off}})$,  red dashed line:  ${\bf B}_{1}(t>T_{\mathrm{on}}) = -{\bf B}_{1}(t<T_{\mathrm{off}})$, blue solid line: magnetically induced nuclear exciton echo with ${\bf B}_{1}(204.3 \mathrm{ns}>t>T_{\mathrm{on}}) = -{\bf B}_{1}(t>204.3 \mathrm{ns}) = -{\bf B}_{1}(t<T_{\mathrm{off}})$. }\label{fig6}
\end{figure}
%%%%%%%%%%%%%%%%%%%%%%%%%%%%%%%%%%%%%%%%%%%%%%%%%%%%%%%%%%%%%%%%%%%%%%%%%
Let us now turn to the measurement of the  $\pi$ phase shift. A typical x-ray optics setup would require to let the $\pi$-modulated photon interfere with a part of the original pulse on a triple Laue interferometer \cite{Laue1,Laue2}.  We  adopt here another approach, namely, the simple and elegant photon echo solution used in NFS experiments with SR \cite{helistoe1982,helistoe1991,Smirnov1996NE,jex1997,Smirnov2005} to allow the scattered photon to interfere with itself in the two-target setup presented in  Fig.~\ref{fig6}(a).   A dynamical magnetic field ${\bf B}_{1}(t)$ is applied to target 1, and a static ${\bf B}_{2}$ is applied to target 2.  The  target response is determined by $R(\alpha,\bigtriangleup_{B},t)=\delta(t)-W(\alpha,\bigtriangleup_{B},t)$ and $W(\alpha,\bigtriangleup_{B},t)=\frac{\alpha}{\sqrt{\alpha\Gamma t}}J_{1}\left( 2\sqrt{\alpha\Gamma t}\right)  e^{-\frac{\Gamma}{2}t+i\bigtriangleup_{B}t}$ \cite{helistoe1991,Roehlsberger2004}, and the  forward-scattered x-ray field  is then given by $E^{(1)}(t)=\int_{0}^{t}R(\alpha,\bigtriangleup_{B},t-\tau)E^{(0)}(\tau)d\tau$ \cite{Smirnov2005}. Using $E^{(0)}(t)=\delta(t)$ as x-ray  input, the resulting electric field  is the real part of
\begin{eqnarray}
\nonumber
E^{(2)}(t) &=& \delta(t)-W(\alpha_{1},\bigtriangleup_{B1},t)-W(\alpha_{2},\bigtriangleup_{B2},t)\\
&+&\int_{0}^{t}W(\alpha_{2},\bigtriangleup_{B2},t-\tau)W(\alpha_{1},\bigtriangleup_{B1},\tau)d\tau.
\label{eq3}
\end{eqnarray}
This depicts the interference of four possible coherent scattering channels \cite{Roehlsberger2004,Smirnov2005}: (1) $\delta(t)$, no scattering; (2)$-W(\alpha_{1},\bigtriangleup_{B1},t)$, the photon is scattered by target 1 only; (3) $-W(\alpha_{2},\bigtriangleup_{B2},t)$, the photon is scattered by target 2 only; (4) the mutual integral, the photon is first scattered by target 1 and then by target 2. Channel (2) and (3)  cancel each other out  when the effective thicknesses of the two targets are equal $\alpha_{1}=\alpha_{2}$ and ${\bf B}_{1}(t>T_{\mathrm{on}})=-{\bf B}_{2}$, i.e,  ${\bf B}_{1}(t)$ is reversed at $t=T_{\mathrm{on}}$. Hence a significant suppression of the  NFS signal can serve as signature for the  effective $\pi$ phase shift  magnetically modulated in target 1. 

In order to obtain the total scattered field intensity, we solve Eqs.~(\ref{eq1}) for both targets using the scattered field of target 1 as incoming field for target 2.  Our numerical results are illustrated in Fig.~\ref{fig6}(b). The presence of two targets results in the faster coherent decay  that proceeds with effective resonant depth of $\alpha=2$, i.e., double the thickness of each target \cite{Smirnov1996NE}.  The magnetic field in target 1 is switched off at $T_{\mathrm{off}}=51$ ns and back on at $T_{\mathrm{on}}=100$ ns.  For continuous phase, the intensity of the scattered field does not change. If, however, the phase of the retrieved field is $\pi$-modulated by turning on the opposite magnetic field $-{\bf B}_1$, the detected signal is  significantly suppressed due to destructive interference between the two scattering channels. In turn,  a second magnetic field rotation back at a node value $E^{(1)}(t>100\  \mathrm{ns})=0$ produces an echo due to constructive interference 
as it can be seen in Fig.~\ref{fig6}(b) for the rotation of ${\bf B}_{1}(t)$ back at $t=204.3$ ns.

This magnetically induced nuclear exciton echo itself provides another convenient solution for  photon storage. A sequence of two $180^{\circ}$ rotations of the magnetic field direction in target 1 at the quantum beat minima can  lead to storage and retrieval of the x-ray photon $\pi$ phase-modulated.  This can be experimentally achieved in antiferromagnets as  $^{57}$FeBO$_{3}$ with strong intrinsic hyperfine magnetic fields that can be rotated with the help of a weak 10 G external field \cite{Shvydko1996}. Fast $180^{\circ}$ magnetic field rotations in such materials  have been demonstrated \cite{Shvydko1994P}. This specific case of magnetic switching in a two-target setup preserves the photon polarization and can modulate the photonic phase but  is less robust compared to our scheme since both efficiency of the storage and the phase of the released photon depend on the rotation moment. Nevertheless, the magnetically induced nuclear exciton echo might provide an additional experimentally accessible setup to investigate mechanical-free x-ray storage and phase modulation of a single-photon wave packet.

In conclusion, 
we have put forward the possibilities of  phase-sensitive storage and $\pi$ phase modulation for single hard x-ray photons in a NFS setup.  These x-ray coherent control tools are important milestones for optics and quantum information applications at shorter wavelengths aiming towards more compact future photonic devices.

We would like to thank  R. R\"ohlsberger for fruitful discussions and T. Herrmannsd\"orfer  for his advice on the generation of strong magnetic fields.

%%%%%%%%%%%%%%%%%%%%%%%%%%%%%%%%%%%%%%%%%%%%%%%%%%%%%%%%%%%%%%%%%%%%%%%%%
\bibliographystyle{unsrt}
\bibliography{NFSBS2A}
%%%%%%%%%%%%%%%%%%%%%%%%%%%%%%%%%%%%%%%%%%%%%%%%%%%%%%%%%%%%%%%%%%%%%%%%%%%%%%%%
%%%%%%%%%%%%%%%%%%%%%%%%%%%%%%%%%%%%%%%%%%%%%%%%%%%%%%%%%%%%%%%%%%%%%%%%%%%%%%%%
\end{document}